\documentclass[twocolumn,twocolappendix]{aastex63}
\usepackage{graphicx}

\shorttitle{Effects of periodic flow on wave propagation}
\shortauthors{Yu}

\begin{document}
\title{Effects of background periodic flow on MHD fast wave propagation to a coronal loop}

\author[0000-0003-1459-3057]{D. J. \surname{Yu}}
\email{djyu79@gmail.com}
\affiliation{Department of Astronomy and Space Science, Kyung Hee University,
 1732, Deogyeong-daero, Yongin, Gyeonggi 17104, Republic of Korea}

\begin{abstract}
We investigate the propagation of MHD fast waves into a cylindrical coronal loop through an inhomogeneous stationary flow region. The background flow is assumed to have a small, spatially periodic structure in addition to a constant speed. We focus on the absorption of the wave energy in Alfv\'{e}n resonance, comparing with the constant flow case. A new flow (absorption) regime is induced by the periodic flow structure which enhances the absorption for the antiparallel flow and inverse absorption (overreflection) for the parallel flow with respect to the axial wave vector, depending on the transitional layer and flow profiles. A giant overreflection and anomalous absorption behavior arise for some flow configurations. In the other flow regimes, its effect on the absorption is shown to be weak.

\end{abstract}


\section{Introduction}
\label{sec1}

Magnetohydrodynamic (MHD) waves in solar atmosphere have been intensively studied for their roles in coronal heating and coronal seismology~\citep{Khomenko2015,Jess2015,Li2020,VanDoorsselaere2020,Banerjee2021,Nakariakov2021}. Coronal mass ejections (CMEs) and flares often generate MHD fast  waves that propagate in all directions~\citep[e.g.,][]{Goedbloed2004,Roberts2019}. Due to this property, MHD fast mode can be a useful tool for the global coronal seismology~\citep{Kwon2013}. The fast waves generated by CME or flare can excite loop oscillations whose property is crucially depending on the topology of the loop location and the wave source~\citep{Selwa2010,Selwa2011,Mandal2021}.

It has been observed that upflows have diverse properties in each local regions of the sun~\citep{Tian2021}. The presence of inhomogeneous flow may enhance the complexity of the wave characteristics, e.g.,~affecting the wave propagation. An intriguing phenomenon is the amplification of wave reflected from the strong velocity shear, so-called overreflection~\citep[see][and references therein]{McKenzie1970,Joarder1997,Mann1999}. {The amplitude of the reflected wave becomes larger than that of the incident wave.} Overreflection is related to the wave that changes its propagation direction (the sign of Doppler-shifted frequency) in the flow region. It requires high velocity shear larger than the phase speed of the wave. The wave with this property is called negative energy wave~\citep[e.g.,][]{Cairns1979,Ryutova1988,Joarder1997,Yu2020}. Shear flow can also enhance or reduce the efficiency of resonant absorption or mode conversion if the shear flow region includes resonance such as Alfv\'{e}n or slow resonance~\citep{Csik1998,Csik2000,Kim2022}. {MHD waves can be unstable near the region with velocity shear and resonance~\citep{Tirry1998,Andries2000}.} \cite{Csik1998} theoretically studied resonant absorption of MHD waves in Alfv\'{e}n and slow resonances where Alfv\'{e}n and cusp speeds have linear profiles in the inhomogeneous region whereas the background flow has step-function profile therein. Their numerical study showed a strong dependence of resonant absorption and overreflection on flow speed.
The wave frequency is also an important factor for these resonance phenomena~\citep{Csik2000}.
In these cases, the wave resonance is involved in overreflection, which needs to be distinguished with the one caused by velocity shear (jump) itself~\citep{Hollweg1990,Andries2001,Kim2022}.

\cite{Kim2022} have recently generalized the model of \cite{Csik1998} by considering both plasma density (Alfv\'{e}n and cusp speeds) and flow speed spatially vary in one direction. A giant overreflection was obtained for some parameter ranges. They interpreted it in terms of cavity-like resonance and the time-reversal-invariant relationship derived in~\cite{Rivero2019}.

It was shown in \cite{Yu2021} that resonant absorption of the coronal loop kink oscillation in Alfv\'{e}n resonance is sensitively dependent on the configurations of the loop and external background flow. In this study, a flow region with a constant flow speed was considered. It was found that the flow region plays the role of potential barrier or well depending on the flow speed.
\cite{Yu2021} argued that the flow region may significantly affect the excitation of the loop oscillations.

This paper considers that the external background flow has a periodic, continuous variation added to a constant speed.
We investigate the effects of the external periodic flows on the propagation of fast waves into an overdense loop and its absorption in the Alfv\'{e}n resonances. The frequency of the incident wave is fixed to that of fundamental kink mode of the loop oscillation. Due to the periodic variation there appears new Alfv\'{e}n resonance in the flow region, which leads to overreflection phenomenon for sufficiently high flow speed.
The model and method are introduced in Sec.~\ref{sec2}, which followed by results in Sec.~\ref{sec3}. We conclude the paper with discussions in Sec.~\ref{sec4}.

\section{Model and Method}
\label{sec2}
As in~\cite{Yu2021}, we consider the propagation of a fast wave through an inhomogeneous flow medium to a coronal loop in cylindrical geometry, for which ideal MHD equations in cold limit is used to derive the governing wave equation. The coronal loop is assumed as a straight and axisymmetric plasma column, which is infinitely long in $z$ direction with radial ($r$) dependence. We assume no radial and azimuthal components for the background magnetic field and flow: $\textbf{\textsl{B}}=(0,0,B_{0})$ and $\textbf{\textsl{U}}=(0,0,U_{0}(r))$. The difference from the previous study is the inclusion of periodic structure in $U_0(r)$ (Eq.~(\ref{eq:3})). The linear wave equation for the total pressure perturbation $P$ can be described as
~\citep[][]{Goossens1992,Yu2021}
\begingroup
\begin{eqnarray}
\frac{d^2 P}{d r^2}+\bigg[\frac{1}{r}-&&\frac{1}{D}\frac{dD}{dr}
\bigg]\frac{d P}{d r}+\bigg[\frac{\Omega^2-\omega_A^2}{v_A^2}-\frac{m^2}{r^2}\bigg]P=0,~~\label{eq:1}
\end{eqnarray}\endgroup
where $D(r)=\rho_0(r)(\Omega^2(r)-\omega_A^2(r))$, $\Omega(r)=\omega-k_zU_0(r)$, $\omega_A^2(r)=k_z^2v_A^2(r)$, $v_A^2(r)=B_{0}^2/\mu_0\rho_0(r)$, $\mu_0$ the magnetic permeability, $\omega$ the wave frequency, $k_z$ the axial wave number, and $m$ the azimuthal wave number.
The density of the coronal loops is assumed to be higher than the outside and varies linearly from $\rho_i$ to $\rho_e$ in the transitional layer:
\begingroup
\begin{eqnarray}
 \rho_0(r)= \left\{ \begin{array}{ll}
         \rho_i & \mbox{ if $r\leq R_0-l$ }\\
         \rho_t(r) & \mbox{ if $R_0-l<r<R_0+l$ }\\
         \rho_e & \mbox{ if  $R_0+l\leq r\leq R$}\\
         \rho_s & \mbox{ if  $r> R$}
        \end{array} \right.,\label{eq:2}
\end{eqnarray}\endgroup where $R=R_0+l+R_u$, $R_0$ the loop radius, $l$ the half thickness of the transitional layer, and $R_u$ the spatial  extent of the shear flow region. The density profile of the transitional layer is given as $\rho_t(r)=(\rho_i-\rho_e)(R_0-r)/2l+(\rho_i+\rho_e)/2$. The density in the wave source region is given as $\rho_s$ which is equal to $\rho_i$ to satisfy the condition that the wave generated at the source region is a propagating mode, $(\omega/k_z)^2>v_{As}^2=B_0^2/(\mu_0\rho_s)$, and also to minimize the interference effects caused by the density difference between $\rho_s$ and $\rho_i$.
 The background flow has a sinusoidal ripple structure outside the loop:
\begingroup
\begin{eqnarray}
 U_0(r)&=&\left\{ \begin{array}{ll}
         0 & \mbox{ if $r< R_0+l$ }\\
         U_e(r) & \mbox{ if $R_0+l\leq r\leq R$ }\\
         0 & \mbox{ if  $r> R$}
        \end{array} \right.,\label{eq:3}
\end{eqnarray}\endgroup
where $U_e(r)=U(1+a\sin(2\pi b(r-R_0-l)/R_u))$ and $U$, $a(>0)$, and $b$ are constants.

{We assume that a fast wave generated outside the inhomogeneous flow region, which surrounds a coronal loop ($r>R$), steadily (continuously) propagates to the loop through the flow region. Our concern is the absorption behavior of the wave energy flux of the incident fast wave through the resonance regions by comparing the incident and scattered fast waves for $r>R$. Our approach describes the steady state picture of the wave propagation behavior including the wave absorption\footnote{It can be thought as large-time asymptotic state of the composite system after a monochromatic cylindrical antenna (with $\omega=\omega_k$) is placed in the source region. For comparing our theoretical approach (IIM) with corresponding numerical simulation, see, e.g.,~\cite{Kim2008} (cf., Fig. 9 therein) and~\cite{Kim2005} .}. }

In the source region ($r>R$) the incident and scattered waves can be described by Bessel ($J_m$) and Hankel ($H_m^{(1)}$) functions of first kind with the scattering coefficient $r_m$~\citep{Yu2016b,Yu2021}:
\begingroup
\begin{eqnarray}
P(r,\phi)&=&\sum_m a_m e^{i m\phi} \big[J_m[k_r(r-R)+c]\nonumber\\
&&+r_m(R)H_m^{(1)}[k_r(r-R)+c]\big],\label{eq:4}
\end{eqnarray}\endgroup
where $k_r(=\sqrt{(\omega^2/v_{As}^2)-k_z^2})$ is the radial wave number for $r>R$, $c$ is a constant equal to $k_rR$, and $a_m$ is an $m$-dependent constant describing the shape of the incident wave. For a plane wave incidence to the $x$ direction, $a_m=i^m$~\citep{Stratton2007}. {We assume that the generated fast wave has only two kink modes ($m=\pm1$).~\citep[][]{Yu2016b,Yu2021}. The value of $a_{\pm1}$ is unimportant here.} The frequency of incident wave is set equal to that of the fundamental standing kink wave, $\omega=\omega_k(=k_zB_0\sqrt{2/(\rho_i+\rho_e)\mu_0})$
, where $k_z=\pi/L$, $L$ is the loop length and ${v_{Ai(e)}=B_0/\sqrt{\mu_0\rho_{i(e)}}}$.

We apply the invariant imbedding method (IIM)~\citep{Klyatskin2005,Yu2016b} to Eq.~(\ref{eq:1}) and Eq.~(\ref{eq:4}), and then obtain a differential equation for $r_m$ \citep{Yu2016b,Yu2021}:
\begingroup\small
\begin{eqnarray}
&&\frac{dr_m(r)}{dr}=\frac{k_r}{H_m}\frac{D(r)}{D_1}[J_m'+r_m(r){H_m^{(1)}}']\nonumber\\
&&+k_r\bigg(\frac{D(r)}{D_1}\frac{{H_m^{(1)}}'}{H_m^{(1)}}+\frac{1}{k_r r}\bigg)
\frac{[J'_m+r_m(r){H_m^{(1)}}'][J_m+r_m(r)H_m^{(1)}]}{J'_mH_m^{(1)}-{H_m^{(1)}}'J_m}\nonumber\\
&&+k_r\bigg(1
-\frac{D_1}{D(r)}\frac{m^2}{(k_rr)^2}
\bigg)\frac{[J_m+r_m(r)H_m^{(1)}]^2}{J'_m H_m^{(1)}-{H_m^{(1)}}' J_m},\label{eq:5}
\end{eqnarray}\endgroup
where $D_1=D(r>R)$, $J_m=J_m(c)$, $H_m^{(1)}=H_m^{(1)}(c)$, and prime means $dz(y)/dy$ for $z(y)$. The initial conditions is $r_m(0)=0$. We integrate Eq.~(\ref{eq:5}) from $0$ to $R$ to calculate $r_m(r=R)$~\citep{Yu2016b,Yu2021}.

When solving Eq.~(\ref{eq:5}), to avoid the singularity due to Alfv\'{e}n resonance ($D(r)=0$), we include a small collision frequency in wave frequency: $\omega\rightarrow\omega+i\omega_i$. To avoid the other singularity which appears at $r=0$ (Eq.~(\ref{eq:5})), we set  the plasma parameters at $r/R_0=\delta$ to those for $r/R_0<\delta$~\citep{Yu2016b,Yu2021}. The values of the two parameters are chosen sufficiently small to not affect the results: $\omega_i=10^{-8}s^{-1}$ and $\delta=10^{-6}$.

From the scattering coefficient $r_m$ we obtain the absorption coefficient $A$
\footnote{When only focusing on the absorption of kink modes, the two kink mode incidence and plane wave incidence have no difference. } as~\citep{Yu2016b}
\begingroup
\begin{eqnarray}
A=-\sum_{m=\pm1}\textmd{Re}(r_m+|r_m|^2).\label{eq:6}
\end{eqnarray}\endgroup
\begin{figure}[]
\includegraphics[width=.4\textwidth]{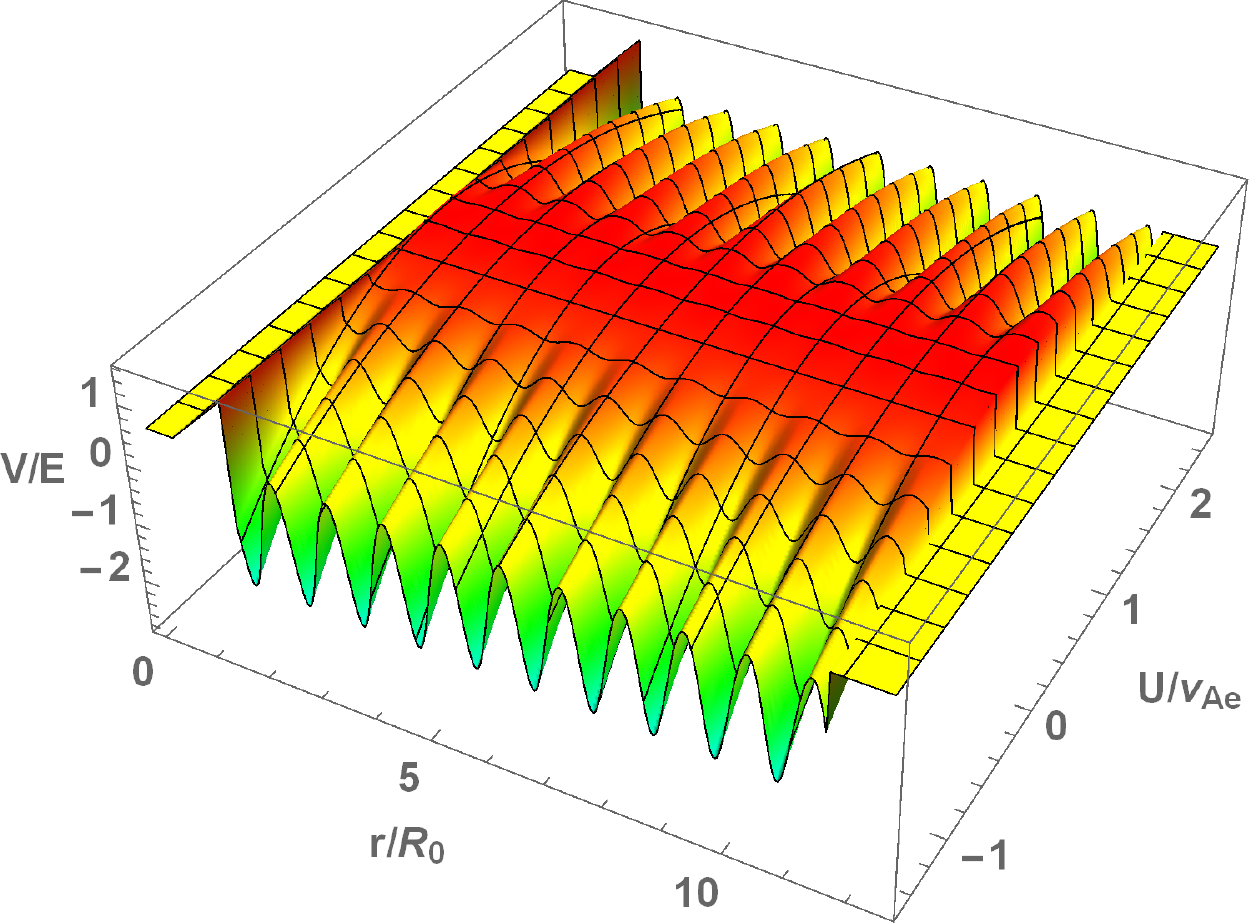}
\caption{\label{f1} A profile of the potential $\delta_{V}=V/E$ as functions of $r/R_0$ and $U/v_{Ae}$ when $\omega=\omega_k$, $L/R_0=50$, $\delta_\rho(=\rho_i/\rho_e)=10$, $l/R_0=0.5$, $R_u/R_0=10$, $a=0.3$, and $b=10$.}
\end{figure}

\begin{figure}[]
\includegraphics[width=0.4\textwidth]{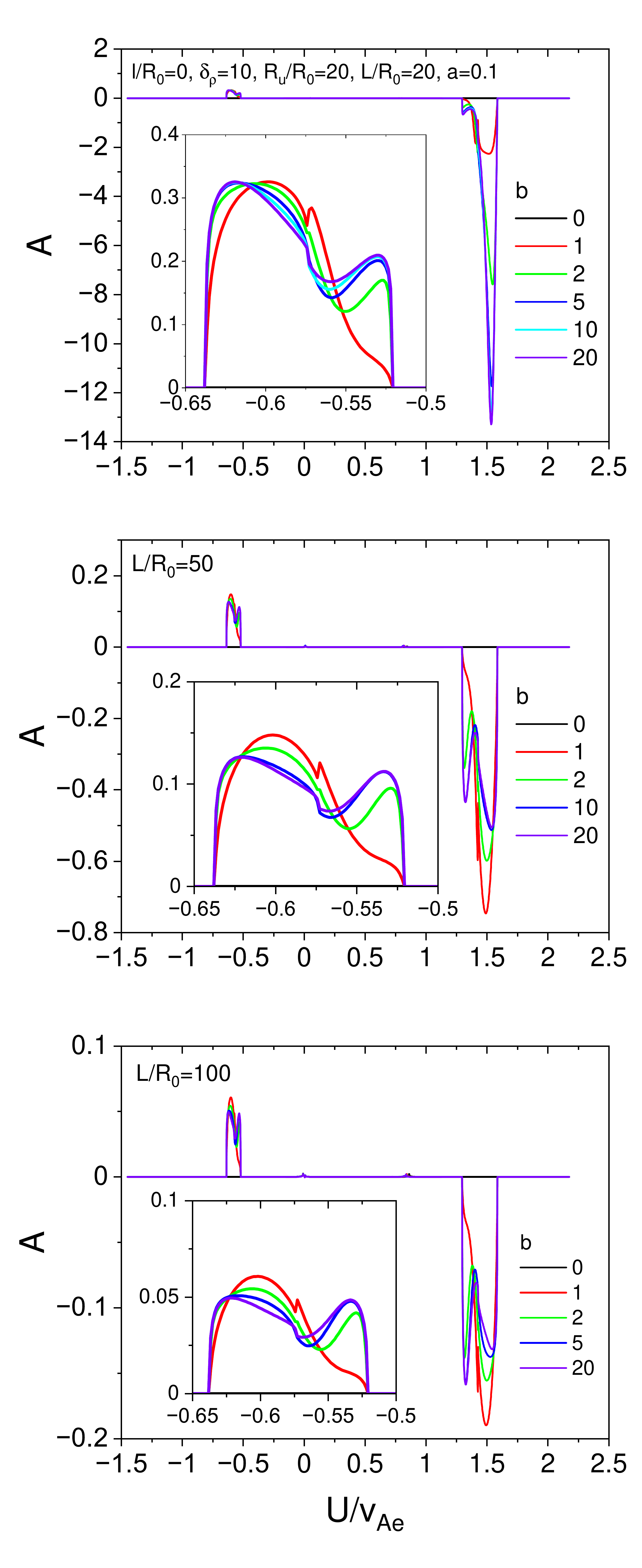}
\caption{\label{f2}Absorption coefficient $A$ vs.~the flow speed ${U/v_{Ae}}$ when $\omega=\omega_k$, ${l/R_0}$=0, $\delta_\rho$=10, $R_u/R_0$=20. From top to bottom, $L/R_0$=20, 50, and 100, and $b$ changes from 0 to 20. Absorption and inverse absorption appear at regime III ($\delta V=1$). In each panel inset shows the absorption behavior in detail.}
\end{figure}
We have previously shown in~\cite{Yu2021} that the potential view is a useful tool to understand the results:
\begin{eqnarray}
\delta_{V}=\frac{V}{E}=1-\bigg[\frac{(\omega-U_0(r)k_z)^2}{v_A^2(r)}-k_z^2\bigg]\frac{v_{As}^2}{\omega^2}.\label{eq:7}
\end{eqnarray}

Fig.~\ref{f1} represents the potential $\delta_{V}$ for the flow model, Eq.~(\ref{eq:3}), as functions of $r/R_0$ and $U/v_{Ae}$ when $\delta_\rho(=\rho_i/\rho_e)$=10, $l/R_0$=0.5, $\omega=\omega_k$, $L/R_0=50$, $R_u/R_0$=10, $a=0.3$, and $b=10$. The previous potential view applies well to the present model. The potential $\delta V$ plays the role of a potential barrier ($\delta_V>1$) or well ($\delta_V<1$), but the symmetry with respect to $U=v_k(=\omega/k_z)$ is now broken due to the periodic ripple. Previously we defined the critical speed $U_c$ as $U_{\pm c}=v_k\pm v_{Ae}$=(-0.5736,1.4264)$v_{Ae}$ for $\delta_{V}=1$ and divided the absorption behaviors into two regimes such that regime I is for $\delta_{V}>1$ ($U_{-c}<U<U_{+c}$) and regime II is for $\delta_{V}<1$ ($U<U_{-c}$ and $U>U_{+c}$)~\citep{Yu2021}. Due to the periodic variation of the background flow, there appear new Alfv\'{e}n resonances in the flow region where the flow speed equals critical speed ($\delta_V=1$). We call this regime III. It is not a point anymore. Its range is determined by $U_e=v_k\pm v_{Ae}$. As $a$ increases, the range of regime III increases, reducing the range of the other regimes. Contrary to the inside of the coronal loop, in flow region there is no constraint to trap the mode converted (Alfv\'{e}n/Alfv\'{e}nic) waves in the Alfv\'{e}n resonance, which freely propagate along the field lines. The appearance of multiple resonances may cause enhanced or decreased absorption. The mode conversion or inverse mode conversion occurs depending on the sign and magnitude of the background flow. If the flow speed $U$ is sufficiently high, the inverse mode conversion could be strong to have $A<0$, so-called over-reflection. In this paper, we call the case for $A<0$ inverse absorption.

\section{Results}
\label{sec3}

To compare with the previous results in ~\cite{Yu2021}, we use the same parameters: $\rho_e=1.67353\times10^{-12}kgm^{-3}$, $B_0=10^{-3}T$, and $R_0=2\times10^6m$, and $v_{Ae}=689.569kms^{-1}$~\footnote{The value of $v_{Ae}$ was incorrectly presented in~\cite{Yu2021}. The contents do not change.}. We first consider the case for the loop with no transitional layer (${l/R_0}=0$).
In Fig.~\ref{f2} we plot the absorption coefficient $A$ vs.~the flow speed ${U/v_{Ae}}$ where $\omega=\omega_k$, $\delta_\rho=10$, ${R}_u/R_0=20$, and ${L/R_0}$ is from 10 to 100. From top to bottom, ${L/R_0}=20$, 50 and 100. The result for $b=0$ implies constant (non-periodic) flow. The absorption appears in regime III ($\delta_V=1$), where its range is proportional to $a|U|$. When $U<0$ $A$ is positive while $A$ is negative when $U>0$. Notice that the range of inverse absorption is larger than that of absorption. For the case $A>0$ ($U<0$),  $A$ is sensitive to the value of $b$, the number of periodicity, and gradually decreases as $L/R_0$ increases. The absorption pattern is similar regardless of the value of $L/R_0$. For the case $A<0$, inverse absorption increases as $b$ increases when $L/R_0$ is small, whereas $A$ has a maximum dip for a small $b$ when $L/R_0$ is large. As $L/R_0$ further increases from 100, $|A|$ gradually decreases and its behavior is similar to the case for $L/R_0=100$.
\begin{figure}[h]
\includegraphics[width=0.48\textwidth]{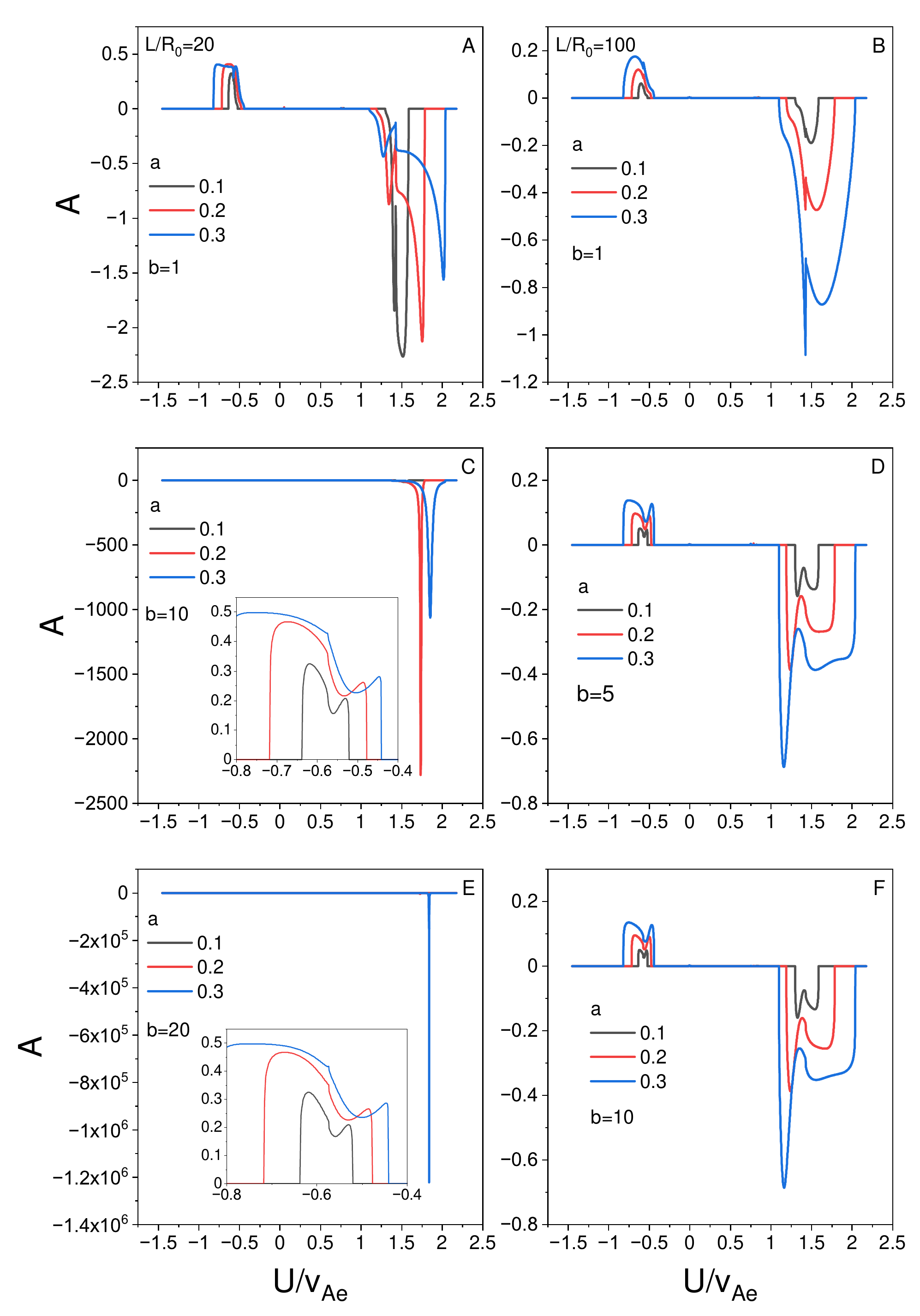}
\caption{\label{f3} $A$ vs. ${U/v_{Ae}}$ for $L/R_0$=20 (A, C, E), 100 (B, D, F) with arbitrary $a$ and $b$. The other parameters are the same as in Fig.~\ref{f2}.}
\end{figure}

\begin{figure}[]
\includegraphics[width=0.45\textwidth]{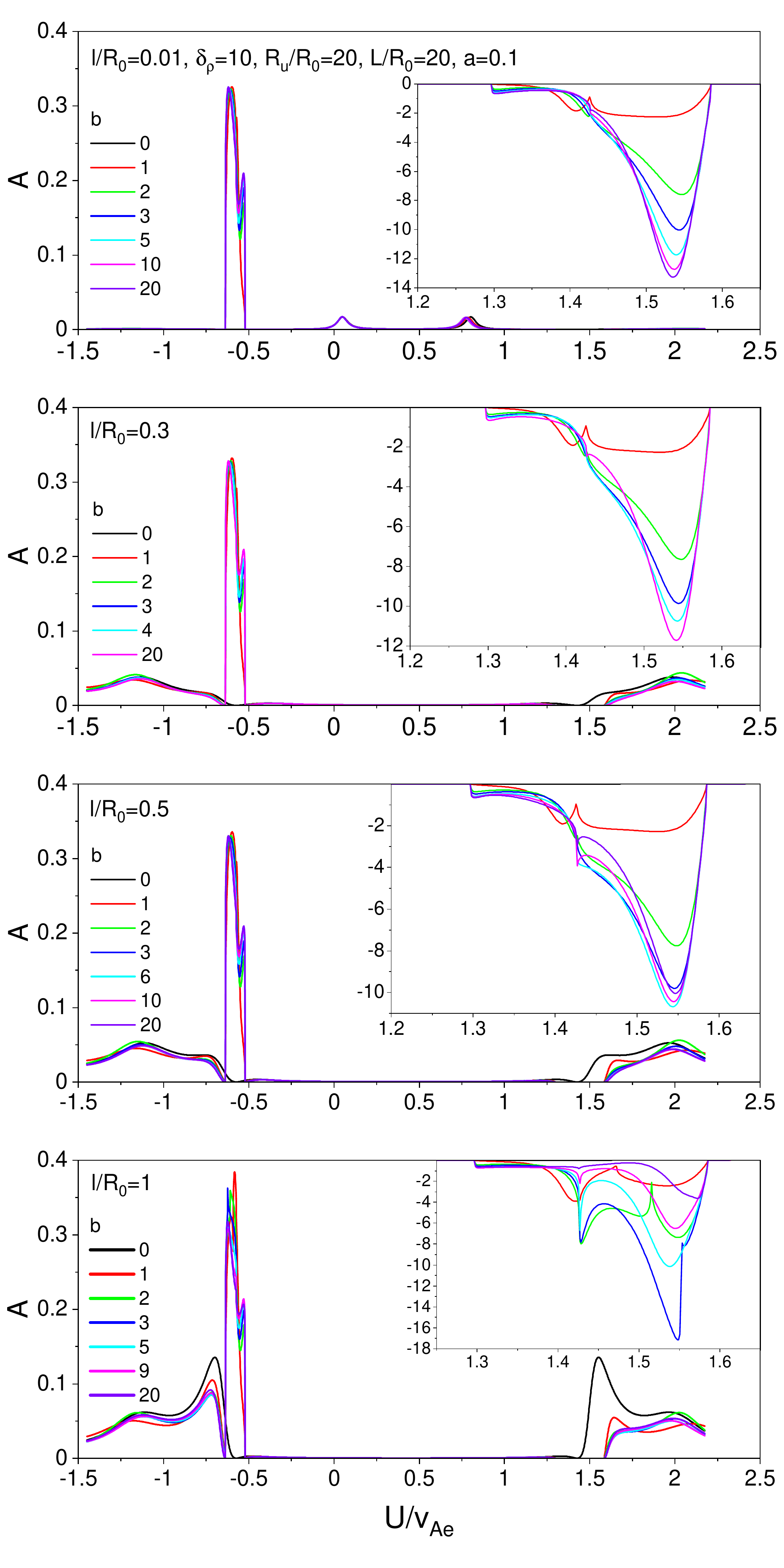}
\caption{\label{f4} $A$ vs. ${U/v_{Ae}}$ for ${L/R_0}$=20 with variation of $l/R_0$ and $b$. The other parameters are the same as in Fig.~\ref{f2}. Each inset shows inverse absorption.}
\end{figure}

Fig.~\ref{f3} presents $A$ vs. ${U/v_{Ae}}$ considering $a$ dependence. From the previous study~\citep{Yu2021} we know that the absorption behavior is different depending on whether the loop length is {short} or long. To compare these typical cases, we choose $L/R_0=$20 (A, C, E) and 100 (B, D, F) and see their dependence on $a$. We first see the right side of $U=0$ $(U>0)$. For $L/R_0=$20 the inverse absorption shows a crucial {dependence} on both $a$ and $b$. It reaches its maximum (local maximum dip) when $a=0.1$ for $b=1$. As $b$ increases the value of $a$ for the maximum dip also increases. On the other hand, for $L/R_0=100$, its has its maximum when $a$ is large regardless of $b$. Another difference is that the inverse absorption is very high for {relatively short} loops. Its value is unreasonably high as shown in panels C and E. Now we see the left side $(U<0)$. For $L/R_0=100$, $\text{max}(A)$ is proportional to $a$ regardless of $b$ and its pattern does not change when $b$ is sufficiently large (D and F). These behavior is similar to $L/R_0=20$ except that $\text{max}(A)\approx0.5$ regardless of $b$ (insets in C, E). The value of $a$ is important for high absorption whereas $b$ is for its complexity.
\begin{figure}[]
\includegraphics[width=0.46\textwidth]{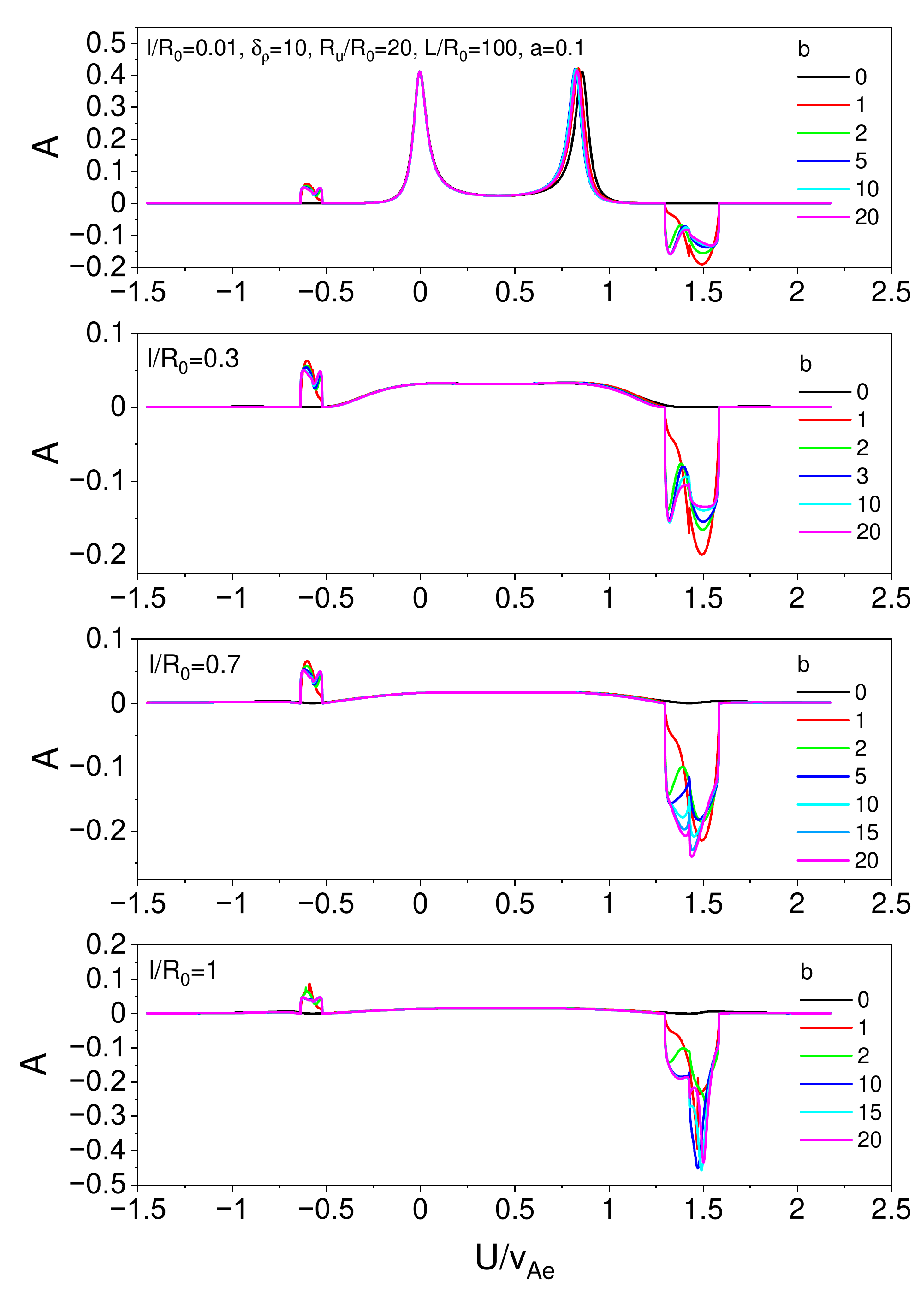}
\caption{\label{f5} Similar to Fig.~\ref{f4} except ${L/R_0}$=100.}
\end{figure}

The presence of transitional layer changes the above results.
Fig.~\ref{f4} shows $A$ vs. $U/v_{Ae}$ for $L/R_0=20$ when $\delta_\rho=10$, $R_u/R_0=20$, and $a=0.1$. Form top to bottom, $l/R_0=$0.01, 0.3, 0.5, and 1. Each inset indicates corresponding inverse absorption and its range. The influence of periodic structure is very small in regime I. A small shift of the second peak to the left occurs (top panel). The influence on absorption in regime II yields reduction of it as $l/R_0$ increases and its effect is more strong in the right side $(U>0)$. On the left side of regime III, $A$ is similar to the case of no transitional layer when $l/R_0$ is small. When $l/R_0$ is close to 1, $A$ is enhanced than $l/R_0=0$ case and shows more complicate aspect depending on $b$. On the right side of regime III (insets), $A$ draws more complicate curves as $l/R_0$ increases. The inverse absorption is also greatest when $l/R_0=1$ as for the absorption of the left side, exceeding non-transitional layer case. The value of $b$ corresponding to the maximum dip depends on $l/R_0$: $b=20$ for $l/R_0=0.01$ and $b=5$ for $l/R_0=1$. The increase of transitional layer enhances the interference effect (see the inset of the bottom panel).

Fig.~\ref{f5} gives $A$ vs. $U/v_{Ae}$ for $L/R_0=100$ when $\delta_\rho=10$, $R_u/R_0=20$, and $a=0.1$.
Form top to bottom, $l/R_0=$0.01, 0.3, 0.7, and 1. The second absorption peak in regime I shows small shift as for $L/R_0=20$ and first peak near $U=0$ does not have noticeable changes since $U$ and $a$ are small.
The absorption and inverse absorption in regime III is not so high as for $L/R_0=20$. In the left side regime $A<0.1$ and in the right side regime $|A|<0.5$. The inverse absorption tends to increases as $l/R_0$.
It is noteworthy that the inverse absorption has a maximum at small $b$ for small $l/R_0$: $b=1$ for $l/R_0=$0.01, 0.3. On the other hand, for large $l/R_0$, large $b$ is needed: $b=20$ for $l/R_0=0.7$ and b=13 for $l/R_0=1$.

\begin{figure*}[]
\includegraphics[width=0.9\textwidth]{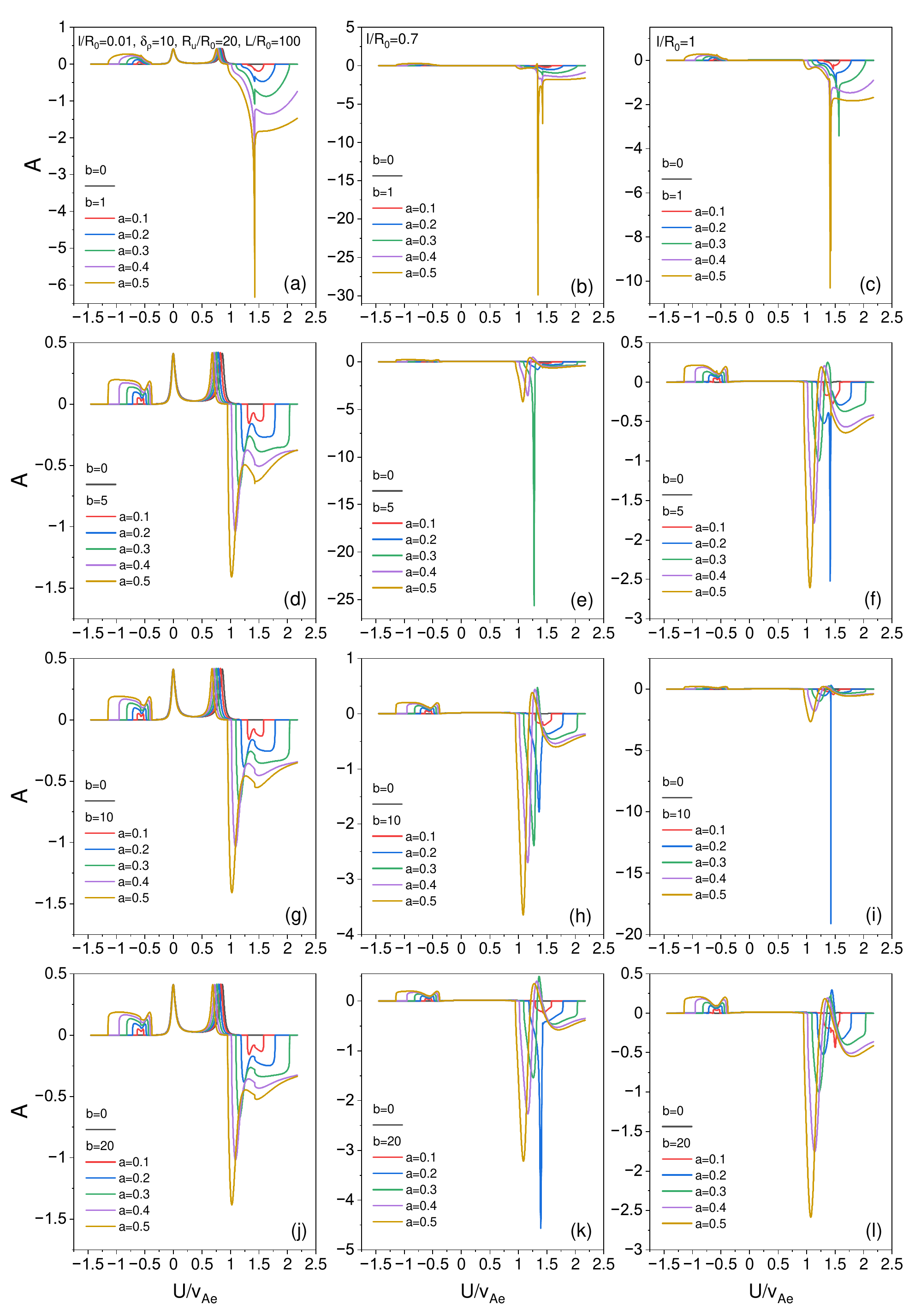}
\caption{\label{f6} Similar to Fig.~\ref{f5} with the variation of $a$ ($L/R_0=100$). From the left column, $l/R_0=0.01, 0.7, 1$. From top to bottom, $b=1,5,10,20$. }
\end{figure*}

Now we consider $a$ dependence of $A$ in the presence of the transitional layer.
Fig.~\ref{f6} describes $A$ vs. $U/v_{Ae}$ by varying $a$ from 0 ($b=0$ case) to 0.5 with 0.1 increment, for $b=1,5,10,20$ when $L/R_0=100$. From the left column $l/R_0=0.01,0.7,1$. When $l/R_0=0.01$ (left column: (a)-(j)), the absorption in regime I has little changes. While the first absorption peak near $U/v_{Ae}=0$ almost does not change its shape and position, the second peak shifts to the left with a small variation as $a$ increases. The value of $b$ has no significant effect on $A$ in this regime. On the other hand, for the inverse absorption in regime III, $A$ is enhanced in proportion to $a$, but its shape nearly does not change when $b>5$. The proportional relationship is also obtained for $b=1$ in top panels: (a)-(c). It is interesting that a position of absorption dip at $U/v_{Ae}\approx1.425$ for $l/R_0=0.01$ and $a=0.5$ diverges into two points when $l/R_0=0.7$, then converges into one point again when $l/R_0=1$. Considering panel (c), the dip position is varying with $a$, which is very different form panel (a). This kind of behavior may be due to the cavity-like resonance in the presence of mode conversion and inverse mode conversion~\citep[e.g.][]{Kim2022}. The strong $a$ dependence of inverse absorption for $b=1$ also implies that the local profile of flow or exactly the form of $d\Omega(r)/dr$ near the resonance point is crucial for the inverse mode conversion, alongside with the density profile of the transitional layer.

The strong interference effect of two opposite resonances is magnificent when $l/R_0$ and $b$ are sufficiently large. It is shown in the rest panels ((e), (f), (h), (i), (k), (l)) that a new absorption regime arises in the inverse absorption regime (the right side of regime III) as $a$ increases. We call this \textit{anomalous absorption}. Comparing with the left side of regime III and with Fig.~\ref{f3} it is clear that the presence of two opposite resonances is essential for this complicate absorption behavior.

\begin{figure*}[]
\includegraphics[width=0.9\textwidth]{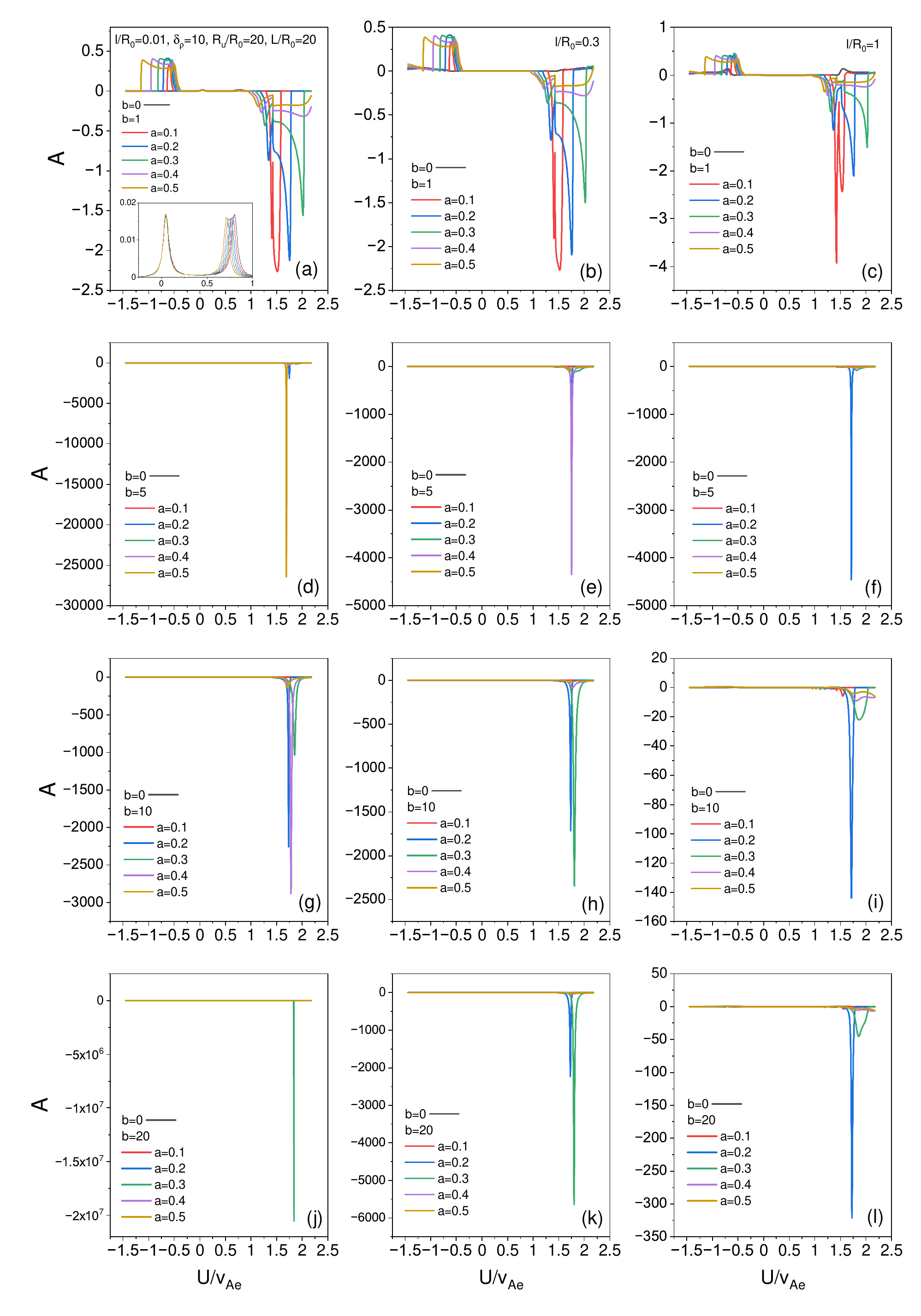}
\caption{\label{f7} Similar to Fig.~\ref{f4} with the variation of $a$ ($L/R_0=20$). From the left column, $l/R_0=0.01, 0.3, 1$. From top to bottom, $b=1,5,10,20$. }
\end{figure*}

The above trend of $A$ changes when the loop length is {short}. We show the results for $L/R_0=20$ in Fig.~\ref{f7}.
For $b=1$ case (top panels: (a)-(c)), $A$ has similarity regardless of $l/R_0$: it has a maximum dip when $a=0.1$, which decreases as $a$ increases. Below the top panels the inverse absorption has strong dependence on $a$ for each $b$. It is remarkable that when (j) $b=20$ and $a=0.3$, $A$ shows a giant overreflection, reaching $\sim-10^7$. On the other hand, the absorption behavior in regime I for $l/R_0=0.01$ (the left column: (a)-(j)) is similar to that of Fig.~\ref{f6} (see the inset of panel (a) in Fig.~\ref{f7}). The increment of transitional layer rapidly reduces the inverse absorption, although it is still strong.

In general, inverse absorption is much stronger for {relatively short} loops with small transitional layer. The anomalous absorption found in Fig.~\ref{f6} is a common property irrespective of $L/R_0$, which is invisible in Fig.~\ref{f7} due to the very high inverse absorption. The (inverse) absorption behaviors in regime III for $l/R_0=0.01$ in Fig.~\ref{f6} and \ref{f7} are very similar to those in Fig.~\ref{f3}.

\section{Conclusions and discussions}
\label{sec4}
We have studied the effects of an intermediate shear flow region, which has a spatial periodic modulation to a constant speed, on the propagation of MHD fast wave into a coronal loop. The wave frequency is fixed to the fundamental kink mode, thus inducing the kink oscillation of the coronal loop and relevant resonant absorption. The presence of sinusoidal periodic variation results in new Alfv\'{e}n resonances in the shear flow region, which induces new flow regime in absorption, regime III ($\delta_V=1$), in addition to regimes I ($\delta_V>1$) and II ($\delta_V<1$) that {was} previously studied in ~\cite{Yu2021}. The incident wave has mode conversion in the Alfv\'{e}n resonance when $\Omega=\omega_{Ae}$ ($U_e=v_k-v_{Ae}$), which yields enhanced absorption. On the other hand, the wave incident on the flow region with $\Omega=-\omega_{Ae}$ ( $U_e=v_k+v_{Ae}$) gains energy from the flow via inverse mode conversion. The reflected wave has a giant amplification for some flow profiles and {relatively short} coronal loops. An anomalous absorption feature is found that an absorption arises in the middle of inverse absorption regime when the transitional layer is sufficiently thick, and the periodicity ($b$) and amplitude of the variation ($a$) are sufficiently large.

The strong amplification of reflection (inverse absorption) for small $b$ may be explained by the cavity-like resonance and the time-reversal-invariant relationship introduced by~\cite{Rivero2019}, who considered the wave propagation through an medium with loss or gain. Using time reversal invariance between two media, they found that a relationship exists among the reflection and transmission coefficients for two media. Their finding means that high absorption in a medium with loss corresponds to high reflection from a medium with gain. In other words, resonant absorption (overreflection) can be treated as the wave absorption (amplification) in a medium with loss (gain). \cite{Kim2022} considered a model that the density linearly varies and the flow speed is constant in the nonuniform layer and verified that there exists a relationship between a near-perfect resonant absorption with a certain flow speed and a giant overreflection with another flow speed. So, there must exists a certain flow speed for the high resonant absorption to obtain a high overrfection (Eq.~(57) therein). A direct application of this argument to our results is not possible since the absorption coefficient $A$ is not simply defined as in \cite{Kim2022} and the flow shear in our model has spatial dependence, but we anticipate that the relevant relationship should exist. When the density and background flow profiles have simple forms, the cavity-like resonance can be a mechanism for strong resonant absorption~\citep[e.g.,][]{Lee1999,Yu2010} and as a result for the giant overreflection. The strong inverse absorption for small $b$ and $l$ seems to be from cavity-like resonance. For the case of large $b$ and $l$, interference effect between scattered waves from multiple resonances has strong influence on the absorption and inverse absorption.

The influence of periodic flow variation on the resonant absorption in regime I is weak. There is a small shift of the peak position and small variation of absorption coefficient as $a$ increases. The change of the first (left) absorption peak at $U\approx0$ is negligibly small compared to the second (right) peak. Its influence on resonant absorption in regime II is negative. It generally reduces the absorption in proportional to $a$. The peak position shifts away from the original position with the increment of $a$ since the regime III extends along with $a$.

Although we have considered a cylindrical plasma the main features of the results may apply to similar situations in plasma slabs. In a slab model an awkward assumption of surrounding shear flow for the loop disappears.

{As shown in~\cite{Csik1998},~\cite{Kim2022} and from the results for $b=1$, a rather simple configuration of shear flow and density can induce overreflection. The inhomogeneity of flow and density (or magnetic field) is the essential ingredient.} Overreflection may be observed in the lower solar atmosphere, solar wind (beyond the Alfv\'{e}nic point), and so on, where the flow shear exceeds the Alfv\'{e}n speed (super-Alfv\'{e}nic) and the phase speed of the wave. If the cusp resonance is of concern, the Alfv\'{e}n speed is replaced with the cusp speed .

{In the phtosphere or chromospher, the flare induced MHD wave may have overrelfection from the sunspot if there is a inhomogenoeus flow structure inside the sunspot or between the flare and a sunspot~\citep[e.g.,][]{Kosovichev2007}. Another object of interest is the coronal hole.~\cite{Zhou2022} have observed a total reflection of flare-driven quasi-periodic extreme ultraviolet wave train at a coronal boundary. To our view, there is also a possible observation of overreflection considering the inhomogeneous structure of outward flows within the coronal hole~\citep{Cranmer2009,Tian2011}. The switch-back with field-aligned flows in solar wind is another possible candidate for the observation of overreflection since the spatially alternating flow and field line can have a periodic potential structure similar to the model considered here~\citep{Chen2021,Neugebauer2021}.}

{If a resonance line for overreflection and another resonance line for resonant absorption reside along a local waveguide structure}, a strong resonant absorption and relevant plasma heating is possible. {Considering the property of omnidirectional propagation of the fast wave, even though there is no waveguide structure, overreflected fast wave can contribute to plasma heating via resonant absorption if the considered plasma structure is sufficiently complicate.} The strong amplification of wave intensity by overreflection may also enhance the nonlinear wave-wave interactions in an inhomogeneous plasma.

{Finally, the inclusion of dissipation mechanisms like, e.g., viscosity may greatly reduce the magnitude of overreflection. Despite this restriction, we anticipate sufficiently high overreflection is to be observable. The study for more realistic situation remains for future work.}

\acknowledgments
The author is grateful to the anonymous referee for fruitful comments.
 This research was supported by Basic Science Research Program through the National Research Foundation of Korea (NRF) funded by the Ministry of Education (No.~2020R1I1A1A01066610, 2021R1I1A1A01045132).

\bibliographystyle{aasjournal}

\end{document}